\documentclass[pra,twocolumn,amsmath,amssymb,amsfonts,showkeys,showpacs,preprintnumbers,floatfix]{revtex4-1}
\usepackage{bm,graphicx,xcolor,subfigure,hyperref,multirow,rotating,lineno}
\newcommand{\eps}{\varepsilon}
\newcommand{\Ec}{E_{\rm c}}
\DeclareMathOperator{\Se}{Se}

\begin{document}
\bibliographystyle{apsrev}

\title{Hooke's law correlation in two-electron systems}

\author{Pierre-Fran\c{c}ois Loos}
\email{loos@rsc.anu.edu.au}
\affiliation{Research School of Chemistry, Australian National University, Canberra, 
Australian Capital Territory 0200, Australia}
\date{\today}

\begin{abstract}
We study the properties of the Hooke's law correlation energy ($\Ec$), 
defined as the correlation energy when two electrons interact {\em via} a harmonic potential 
in a $D$-dimensional space. 
More precisely, we investigate the $^1S$ ground state properties of two model systems: 
the Moshinsky atom (in which the electrons move in a quadratic potential) and the spherium model 
(in which they move on the surface of a sphere). 
A comparison with their Coulombic counterparts is made, 
which highlights the main differences of the $\Ec$ in both the weakly and strongly correlated limits.  
Moreover, we show that the Schr\"odinger equation of the spherium model is exactly solvable 
for two values of the dimension ($D = 1 \text{ and } 3$), 
and that the exact wave function is based on Mathieu functions.
\end{abstract}


\maketitle

\section{\label{sec:intro}Introduction}

Understanding correlation effects remains the central problem 
in theoretical quantum chemistry and physics, 
and the main goal of most of the new theories and models in this research area \cite{Helgaker,ParrYang}.  
In order to gain some insight into electron correlation, 
two-electron model systems have always played a key role 
by shedding new light on the relative motion of electrons.

Two such models are the Hooke's law atom (or harmonium, or hookium) 
\cite{Kestner62,Kais89,Taut93,Cioslowski00} 
and the spherium model \cite{Ezra82,Ezra83,Ojha87,Hinde90,Warner85,Seidl07b,TEOAS1,Quasi09}.  
In hookium, the two electrons are bound to the nucleus by a harmonic potential, 
while in spherium, 
the position of the electrons are restricted to remain on the surface of a sphere. 
In both cases, the electrons repel Coulombically.
For these systems, the exact solution of the Schr\"odinger equation can be obtained 
for some discrete values of the confinement parameter 
(the force constant for hookium and the radius of the sphere for spherium) \cite{Kais89,Taut93,Quasi09}.  
Consequently, these model systems (and others \cite{Alavi00,Jung03,Thompson02,Thompson04,Thompson05}) 
have been extensively used to test various approximations within 
density functional theory (DFT) \cite{Filippi94,Taut98,Ivanov99,Seidl00,Seidl07b,Sun09,GoriGiorgi09a,Seidl10}.
However, the Hartree-Fock (HF) solution is not always available in closed-form; in hookium for example, 
although accurate solutions have been found \cite{ONeill03,Ragot08}, 
no closed-form expression of the ground state HF orbital has been obtained yet.

According to L\"owdin \cite{Lowdin59}, the correlation energy ($\Ec$) is defined as the error
\begin{equation} \label{Ec-def}
	\Ec = E - E_{\rm HF},
\end{equation}
which pertains to the HF approximation \cite{Deng10}.
$\Ec$ is a function of the external potential $V(r)$, the dimensionality of the space where the 
electrons are moving ($D$), and the interelectronic potential $w(u)$, 
where $u \equiv | \bm{r}_1 - \bm{r}_2 |$ is the distance between the two electrons.

We define the Coulombic and the Hooke's law correlation energies
as the correlation energies when the two electrons interact {\em via} 
a repulsive Coulomb potential [$w(u) = u^{-1}$] or an attractive harmonic potential 
[$w(u) = \omega^2 u^2/2$], respectively.

We have recently shown that the Coulombic correlation energy is rather 
insensitive to $V(r)$ (at least in the high-density limit \cite{EcLimit09}), 
but strongly dependent on the number of degrees of freedom of the electron pair ($D$).  
However, the question of how the interelectronic potential $w(u)$ influences the correlation 
energy has not yet been addressed, and this is indeed the purpose of this article.

In the following, we will consider the $^1S$ ground state of the Hooke's law analog of $D$-hookium 
(known as the Moshinsky atom \cite{Moshinsky68}), and $D$-spherium (labeled as $D$-HL-spherium in the following).
We adopt the convention that a $D$-sphere is the surface of a ($D+1$)-dimensional ball.  
Thus, in 2-spherium, for example, this is the surface of a three-dimensional ball.  
For $D = 3$, both the exact and HF solutions of the Moshinsky atom are known \cite{Moshinsky68}.  
We will generalize these results for any value of $D$ in Section \ref{sec:Moshinsky}, 
and derive the two-electron probability distributions in both the position and momentum space.
In Section \ref{sec:TEOAS}, we analyze the energy behavior of $D$-HL-spherium in both the 
weakly and strongly correlated regime.  
Moreover, we demonstrate that the exact solution of the Schr\"odinger equation 
can be found for two values of the dimensionality ($D = 1$ and $D = 3$).  
Atomic units are used throughout.

\section{\label{sec:Moshinsky}Moshinsky atom}

\begin{figure}                  
\begin{center}
\caption{\label{fig:Ec}$-\frac{\Ec}{D}$ in the Moshinsky atom as a function of $\omega$ (solid).  
The small-$\omega$ (dashed) and large-$\omega$ (dotted) expansion are also represented.}
        \includegraphics[width=0.45\textwidth]{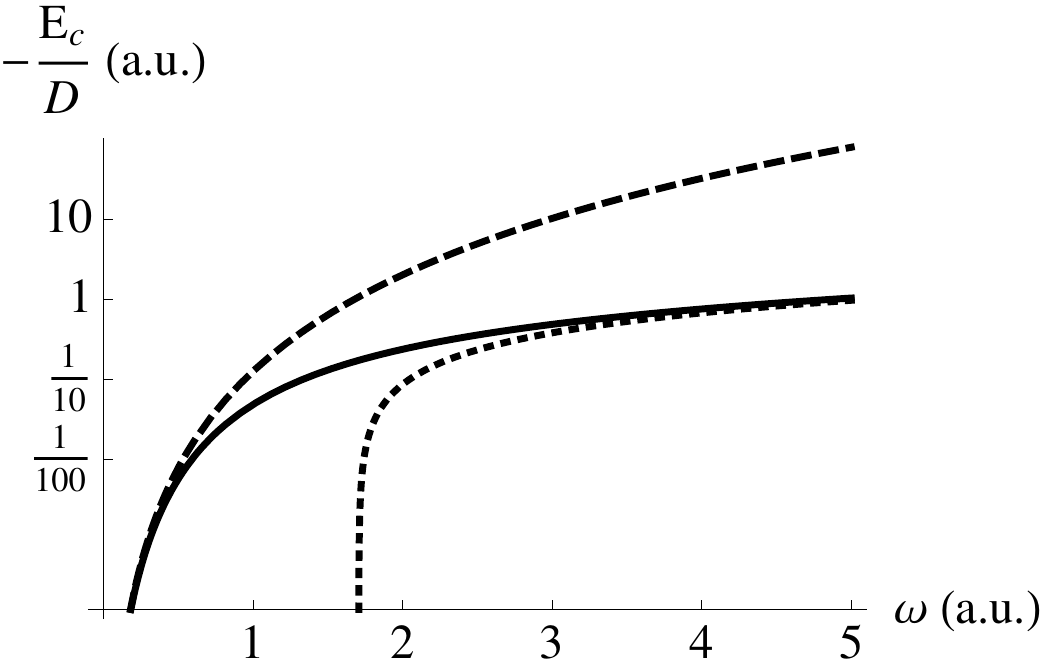}
\end{center}
\end{figure}

\subsection{\label{subsec:Moshinsky-exact}Exact solution}

The Moshinsky atom \cite{Moshinsky68} is defined by the Hamiltonian
\begin{equation} \label{H}
	\Hat{H}		= 	- \frac{1}{2} \left( \nabla_{\bm{r}_1}^2 
				+ \nabla_{\bm{r}_2}^2 \right) 
				+ \frac{1}{2} \left( r_1^2 + r_2^2 \right) 
				+ \frac{\omega^2}{2} \left| \bm{r}_1 - \bm{r}_2 \right|^2,
\end{equation}
where $\omega^2$ is the force constant between the two electrons.  
The Hamiltonian \eqref{H} can be separated into the extracule and intracule coordinates, 
which reads respectively
\begin{align}
	\bm{\Lambda}	&	= \frac{\bm{r}_1 + \bm{r}_2}{\sqrt{2}},		&	
	\bm{\lambda}	&	= \frac{\bm{r}_1 - \bm{r}_2}{\sqrt{2}},
\end{align}
yielding
\begin{equation}
	\Hat{H} 	= 	\frac{1}{2} \left[ - \nabla_{\bm{\Lambda}}^2 + \Lambda^2 \right] 
				+ \frac{1}{2} \left[ - \nabla_{\bm{\lambda}}^2 + (2\omega^2+1) \lambda^2 \right].
\end{equation}
For $S$ states in a $D$-dimensional space, the Laplace operator is given by \cite{Herschbach86}
\begin{equation}
	\nabla_{\bm{r}}^2 = \frac{\partial^2}{\partial r^2} + \frac{D-1}{r} \frac{\partial}{\partial r}.
\end{equation}
This leads to the exact wave function
\begin{equation} \label{psi}
	\psi (\bm{\Lambda},\bm{\lambda})	
	= \frac{(2\omega^2+1)^{D/8}}{\pi^{D/2}} 
	e^{-\frac{1}{2} \Lambda^2} e^{-\frac{1}{2} \sqrt{2\omega^2+1} \lambda^2},
\end{equation}
and energy
\begin{equation} \label{E}
	E = \frac{D}{2} \left( 1 + \sqrt{2\omega^2+1} \right).
\end{equation}
Eq. \eqref{E} reveals that the exact energy is linear
with respect to the dimensionality $D$, due to the separability of the Hamiltonian \eqref{H}.

\subsection{\label{subsec:Moshinsky-HF}Hartree-Fock approximation}

According to the Hartree-Fock (HF) approximation \cite{Szabo}, the total HF wave function of the singlet state 
is defined as 
\begin{equation} \label{psi-HF}
        \psi_{\rm HF} (\bm{r}_1,\bm{r}_2) = 	\varphi_{\rm HF} \left( \bm{r}_1 \right) 
						\varphi_{\rm HF} \left( \bm{r}_2 \right),
\end{equation}
where $\varphi_{\rm HF}$ is the HF orbital, eigenfunction of the Fock operator
\begin{equation}
        \Hat{F} 	= 	- \frac{1}{2} \nabla_{\bm{r}}^2 
				+ \frac{1}{2} r^2 
				+ \frac{\omega^2}{2} 
				\int_{\bm{r}_2} \varphi_{\rm HF}^2 \left( \bm{r}_2 \right) r_2^2\,d\bm{r}_2,
\end{equation}
associated with the eigenvalue
\begin{equation} \label{eps-HF}
	\eps_{\rm HF} = \frac{D}{4} \frac{3\omega^2+2}{\sqrt{\omega^2+1}}.
\end{equation}
It can be shown that 
\begin{equation}
	\varphi_{\rm HF} (\bm{r}) = \frac{(\omega^2+1)^{D/8} }{\pi^{D/4}} 
				e^{-\frac{1}{2} \sqrt{\omega^2+1} r^2}.
\end{equation}
Then, \eqref{psi-HF} is easily recast as
\begin{equation}
        \psi_{\rm HF} (\bm{\Lambda},\bm{\lambda}) 
	= \frac{(\omega^2+1)^{D/4}}{\pi^{D/2}} 
	e^{-\frac{1}{2} \sqrt{\omega^2+1} \left(\Lambda^2+\lambda^2\right)}.
\end{equation}
The total HF energy, which also behaves linearly with $D$, is
\begin{equation} \label{E-HF}
	E_{\rm HF} = D \sqrt{1+\omega^2}.
\end{equation}

\subsection{\label{subsec:Moshinsky-Ec}Correlation energy}

According to \eqref{Ec-def}, the explicit expression of the correlation energy of the Moshinsky atom reads
\begin{equation} \label{Ec}
	\Ec = \frac{D}{2} \left(1- 2\sqrt{\omega^2+1} + \sqrt{2\omega^2+1}\right),
\end{equation}
which obviously decreases linearly with $D$.

Eqs. \eqref{E}, \eqref{E-HF} and \eqref{Ec} yield the small-$\omega$ expansion
\begin{align}
	\frac{E}{D} 		&	= 	1 	+ 	\frac{\omega^2}{2} 	
							- 	\frac{\omega^4}{4}	
							+	O\left(\omega^6\right),	\\
	\frac{E_{\rm HF}}{D} 	&	= 	1	+	\frac{\omega^2}{2}	
							-	\frac{\omega^4}{8}	
							+	O\left(\omega^6\right),	\\
	\frac{\Ec}{D} 		&	= 		-	\frac{\omega^4}{8}	
							+	O\left(\omega^6\right),	\label{Ec-small-w}
\end{align}
and the large-$\omega$ expansion
\begin{align}
	\frac{E}{D} 		&	= 	\frac{\omega}{\sqrt{2}}				
					+	\frac{1}{2}		
					+	\frac{1}{4\sqrt{2}\omega}
					+	O\left(\frac{1}{\omega^2}\right),	\\
	\frac{E_{\rm HF}}{D} 	&	= 	\omega	
					+	\frac{1}{2\omega}
					+	O\left(\frac{1}{\omega^2}\right),		\\
	\frac{\Ec}{D} 		&	= 	\left(\frac{1}{\sqrt{2}}-1\right) \omega	
					+	\frac{1}{2}		
					+	\frac{\sqrt{2}-4}{8\omega}
					+	O\left(\frac{1}{\omega^2}\right).	\label{Ec-large-w}
\end{align}

The correlation energy of the Moshinsky atom, represented in Figure \ref{fig:Ec}, 
is quartic for small $\omega$ (weakly correlated regime), 
and it decreases as $\omega$ in the large-$\omega$ regime (strongly correlated regime).
In comparison, the correlation energy of $D$-hookium in the weakly correlated limit 
tends to a constant \cite{Cioslowski00,EcLimit09}.

In the strongly correlated regime ($\omega \to \infty$), 
the exact and HF wave functions of the Moshinsky atom become
\begin{align}
	\psi (\bm{\Lambda},\bm{\lambda}) 	& = \frac{\left(2 \omega \right)^{D/8}}{\pi^{D/2}}
						e^{-\frac{1}{2}\Lambda^2}	
						e^{-\frac{\omega}{\sqrt{2}}\lambda^2},		\\
	\psi_{\rm HF} (\bm{\Lambda},\bm{\lambda}) 	& = \frac{\omega^{D/2}}{\pi^{D/2}} 
							e^{-\frac{1}{2} \omega 
							\left(\Lambda^2+\lambda^2\right)},
\end{align}
associated with the energies
\begin{align}
	\frac{E}{D} 		&	= \frac{\omega}{\sqrt{2}} + \frac{1}{2}, \label{asy-Moshinsky}	\\
	\frac{E_{\rm HF}}{D} 	&	= \omega + \frac{1}{2\omega}.	\label{asy-Moshinsky-HF}  
\end{align}
In \eqref{asy-Moshinsky}, the second term ($1/2$) is related with the motion of the center of mass, 
while the first term ($\omega/\sqrt{2}$) is associated with the zero-point oscillations of the electrons. 
Indeed, in the strongly correlated regime, the electrons oscillate 
with an angular frequency $D\sqrt{2}\omega$.
Such phenomena are ubiquitous in strongly correlated systems,
as demonstrated by Gori-Giorgi, Seidl and their coworkers
\cite{Seidl00,Seidl99a,Seidl99b,Seidl07a,Seidl07b,GoriGiorgi09b,GoriGiorgi09c,Seidl10}􏶬.

As one can see, the HF solution does not describe properly these oscillations, 
and thus exhibits a wrong behavior in the large-$\omega$ limit.

\subsection{\label{subsec:P}Position Intracule}

\begin{figure}
\begin{center}
\caption{\label{fig:P}$\mathcal{P}$ (solid), $\mathcal{P}_{\rm HF}$ (dashed) and $\Delta\mathcal{P}$ (dotted) 
in the Moshinsky atom as a function of $u$ for the unit force constant and $D=3$}
	\includegraphics[width=0.45\textwidth]{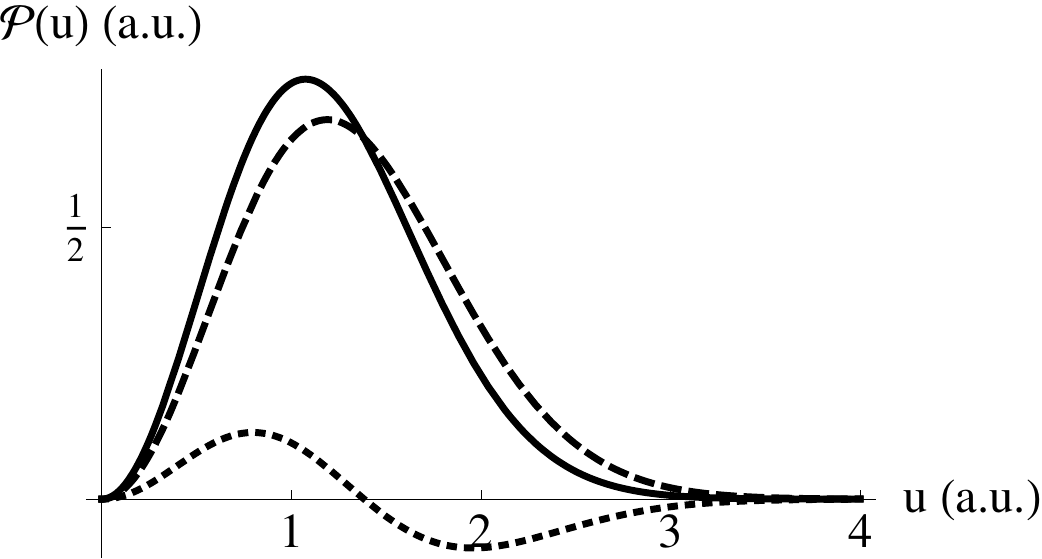}
\end{center}
\end{figure}

To study further the correlation effects in Hooke's law systems, we have determined the position intracule 
$\mathcal{P}(u)$ of the Moshinsky atom ($D \ge 2$). 
$\mathcal{P}(u)$ gives the probability of finding two electrons separated by a distance $u$ 
\cite{Coulson61,Gill00}, and is defined by
\begin{equation}
	\mathcal{P}(u) 
	= \int_{(\bm{\Omega}_u)} 
	\psi (\bm{r}_1,\bm{r}_2)^2 \delta(\left| \bm{r}_{1} - \bm{r}_{2} \right| - u)\,
	d\bm{\Omega}_u,
\end{equation}
where $\bm{\Omega}_u$ is the angular component of $\bm{u}$, 
and $\delta$ the Dirac delta function \cite{ASbook}.
From \eqref{psi} and \eqref{psi-HF}, it follows
\begin{align}
	\mathcal{P}(u,\omega) 		& 	= 	\frac{(2\omega^2+1)^{D/4} u^{D-1}}
							{2^{\frac{D}{2}-1} \Gamma \left(\frac{D}{2}\right)} 
							e^{-\frac{1}{2} \sqrt{2\omega^2+1} u^2},	\\
	\mathcal{P}_{\rm HF}(u,\omega)	&	= 	\frac{(\omega^2+1)^{D/4} u^{D-1}}
							{2^{\frac{D}{2}-1} \Gamma \left(\frac{D}{2}\right)} 
							e^{-\frac{1}{2} \sqrt{\omega^2+1} u^2}.
\end{align}
where $\Gamma$ is the gamma function \cite{ASbook}. 
Surprisingly, we have the relation
\begin{equation} \label{P-PHF}
	\mathcal{P}_{\rm HF}(u,\omega) = \mathcal{P}(u,\frac{\omega}{\sqrt{2}}),
\end{equation}
which means the HF intracule is related to the exact one by a scaling of the confinement strength.
This kind of relation could be useful in the future development of intracule functional theory 
\cite{Gill05,Dumont07,Crittenden07a,Crittenden07b,Bernard08,Pearson09b}.

In analogy with the Coulomb hole \cite{Coulson61}, 
we define the Hooke hole as the difference between the position intracule obtained 
from the exact wavefunction and the corresponding HF one:
\begin{equation}
        \Delta \mathcal{P}(u,\omega) = \mathcal{P}(u,\omega) - \mathcal{P}_{\rm HF}(u,\omega).
\end{equation} 

Contrary to the Coulomb correlation \cite{Coulson61,Pearson09a}, in the Moshinsky atom
the correlation increases the likelihood of finding the two electrons close together 
and decreases the probability of larger values of $u$.
It implies that the Hooke hole is positive for small $u$ and negative for larger $u$. 
This behavior is due to the attractive nature of the harmonic potential between the electrons.
It is illustrated in Figure \ref{fig:P}, where we have reported the 
position intracules and the Hooke hole of the Moshinsky atom for $\omega = 1$ and $D=3$.

\subsection{\label{subsec:M}Momentum Intracule}

\begin{figure}
\begin{center}
\caption{\label{fig:M}$\mathcal{M}$ (solid), $\mathcal{M}_{\rm HF}$ (dashed) and $\Delta\mathcal{M}$ (dotted) 
in the Moshinsky atom as a function of $v$ for the unit force constant and $D=3$.}
	\includegraphics[width=0.45\textwidth]{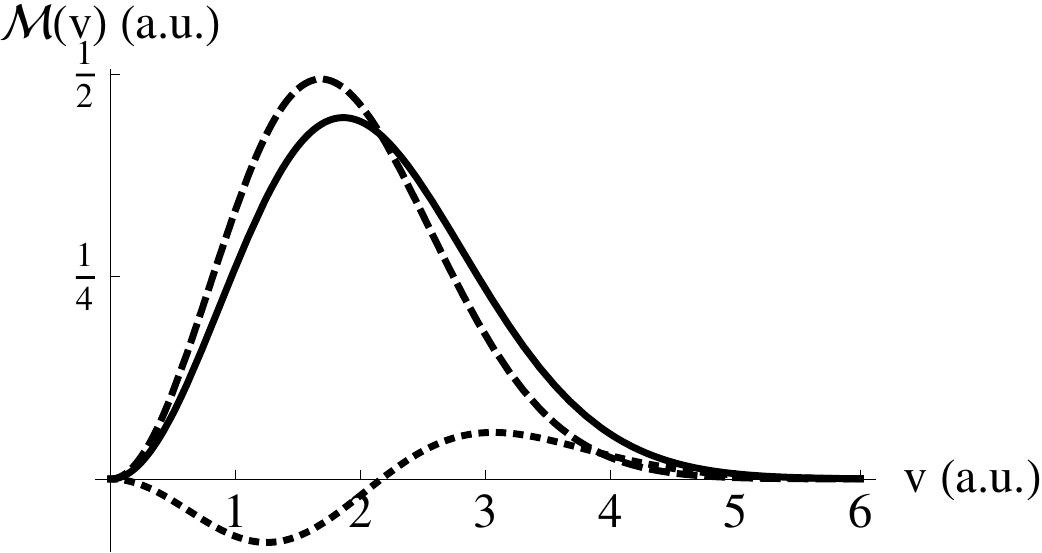}
\end{center}
\end{figure}

The momentum intracule \cite{Besley02}, 
which gives the probability of finding the two electrons moving with a relative momenta $v$,
has been computed to get some insight into the relative momenta of the electrons:
\begin{equation}
	\mathcal{M}(v) 
	= \int_{(\bm{\Omega}_v)} 
	\phi (\bm{p}_1,\bm{p}_2)^2 \delta(\left| \bm{p}_{1} - \bm{p}_{2} \right| - v)\,
	d\bm{\Omega}_v,
\end{equation}
where $\phi(\bm{p}_1,\bm{p}_2)$ is the momentum wave function, 
$\bm{p}_1$ and $\bm{p}_2$ are the momenta of electrons 1 and 2, 
and $\bm{\Omega}_v$ is the angular component of $\bm{v}$.
$\phi(\bm{p}_1,\bm{p}_2)$ is obtained from the position wave function by a Fourier transform:

\begin{align}
	\phi (\Bar{\bm{\Lambda}},\Bar{\bm{\lambda}})	
		& 	= 
			\frac{\left(2\omega^2+1\right)^{-D/8}}{\pi^{D/2}}
			e^{-\frac{1}{2}\Bar{\Lambda}^2}
			e^{-\frac{1}{2\sqrt{2\omega^2+1}}\Bar{\lambda}^2},	
			\label{phi-exact}\\
	\phi_{\rm HF} (\Bar{\bm{\Lambda}},\Bar{\bm{\lambda}})
		& 	= 
			\frac{\left(\omega^2+1\right)^{-D/4}}{\pi^{D/2}}
			e^{-\frac{1}{2\sqrt{\omega^2+1}}\left(\Bar{\Lambda}^2+\Bar{\lambda}^2\right)},
			\label{phi-HF}
\end{align}
where 
\begin{align}
        \Bar{\bm{\Lambda}}    &       = \frac{\bm{p}_1 + \bm{p}_2}{\sqrt{2}},         &       
        \Bar{\bm{\lambda}}    &       = \frac{\bm{p}_1 - \bm{p}_2}{\sqrt{2}},
\end{align}
are the extracule and intracule coordinates in momentum space.
Using \eqref{phi-exact} and \eqref{phi-HF}, the exact and HF momentum intracules are given by
\begin{align}
	\mathcal{M} (v,\omega)		& = \frac{\left(2\omega^2+1\right)^{-D/4} v^{D-1}}
					{2^{\frac{D}{2}-1}\Gamma\left(\frac{D}{2}\right)}
					e^{-\frac{1}{2\sqrt{2\omega^2+1}} v^2},	
					\label{Mexact}\\
	\mathcal{M}_{\rm HF} (v,\omega)	& = \frac{\left(\omega^2+1\right)^{-D/4} v^{D-1}}
					{2^{\frac{D}{2}-1}\Gamma\left(\frac{D}{2}\right)}
					e^{-\frac{1}{2\sqrt{\omega^2+1}} v^2}.
					\label{MHF}
\end{align}
From \eqref{Mexact} and \eqref{MHF}, we show that the property \eqref{P-PHF} related to the position intracule 
is still valid for the momentum intracule:
\begin{equation}
	\mathcal{M}_{\rm HF}(v,\omega) = \mathcal{M}(v,\frac{\omega}{\sqrt{2}}).
\end{equation}
The results are gathered in Figure \ref{fig:M}, 
where we have represented the momentum intracules and the Hooke hole in momentum space
\begin{equation}
        \Delta \mathcal{M}(v,\omega) = \mathcal{M} (v,\omega) - \mathcal{M}_{\rm HF} (v,\omega).
\end{equation}
The exact and HF momentum intracules exhibit a similar shape to the position intracules, but
the Hooke hole in momentum space is drastically different from its position space counterpart. 
Indeed, the correlation in momentum space favors electrons moving with high relative momentum, 
while correlation in position space evidences a decrease of the probability 
of finding electrons further apart.
Once again, this behavior is different in hookium, 
in which correlation favors both lower and higher relative momenta \cite{ONeill03}.
This is also due to the attractive nature of the interelectronic potential.

\section{\label{sec:TEOAS}Hooke's law spherium}

\subsection{\label{subsec:TEOAS-Exact}Expansion of the exact energy}

In terms of the interelectronic angle ($0 \le \theta \le \pi$), 
the Hamiltonian of two electrons on a $D$-sphere of radius $R$, 
and interacting with a force constant $\Omega^2$ is \cite{Seidl07b,TEOAS1,Quasi09}
\begin{equation} \label{H-TEOAS}
	\Hat{H}		= 	- \frac{1}{R^2} \left[ \frac{d^2}{d\theta^2} 	
				+ \left(D-1\right) \cot \theta \frac{d}{d\theta} \right] 
				+ \Omega^2 R^2 \left(1 - \cos \theta \right).
\end{equation}
After the energy scaling $E \leftarrow R^2 E$, and the definition 
of the dimensionless variable $\omega = \Omega R^2$, 
The Hamiltonian \eqref{H-TEOAS} reduces to
\begin{equation} \label{Hs-TEOAS}
	\Hat{H}		= 	- \left[ \frac{d^2}{d\theta^2} 	
				+ \left(D-1\right) \cot \theta \frac{d}{d\theta} \right] 
				+ \omega^2 \left(1 - \cos \theta \right).
\end{equation}

Then, perturbation theory \cite{White70,Cioslowski00,HookCorr05,Seidl07b,TEOAS1,EcLimit09} 
can be applied, and 
expanding $E$ up to the second-order in $\omega^2$, it is straightforward to show that
\begin{equation} \label{smallR-TEOAS}
	E	\simeq	\omega^2 - \frac{\omega^4}{D(D+1)} + \ldots,
\end{equation}
which is valid for the small-$\omega$ regime (weakly correlated limit).

For large $\omega$ (strongly correlated limit), 
the potential dominates the kinetic energy and the electrons tend to 
localize on the same side of the sphere, and oscillate around 
their equilibrium position (zero-point oscillations).
However, in Coulombic systems, the localization takes place on opposite sides of the sphere 
(formation of a Wigner molecule \cite{Wigner34}), because of the repulsive nature of the Coulombic interaction.

In this limit, the Hamiltonian \eqref{Hs-TEOAS} becomes, for small oscillations ($\theta \simeq 0$):
\begin{equation} \label{H0}
	\Hat{H}^{(0)}	= 	- \left[ \frac{d^2}{d\theta^2} 
			+ 	\frac{D-1}{\theta} \frac{d}{d\theta} \right] 
			+ 	\frac{\omega^2}{2} \theta^2,
\end{equation}
where we have used the Taylor expansions:
\begin{align}
	\cot \theta 	&	\simeq 	\frac{1}{\theta} - \frac{\theta}{3} + \ldots,		&
	\cos \theta 	&	\simeq 	1 - \frac{\theta^2}{2} + \ldots
\end{align}
The corresponding ground state eigenfunction and eigenvalue of \eqref{H0} are
\begin{align} \label{psi-largeR}
	\psi^{(0)} (\theta) 	& 	= \sqrt{\frac{1}{2^{\frac{D}{4}-1} \Gamma \left(\frac{D}{2}\right)}} 
					e^{- \frac{\omega}{2\sqrt{2}} \theta^2},		&
	E^{(0)}			&	= \frac{D\omega}{\sqrt{2}}.
\end{align}
The electrons are localized on the same side of the sphere, and oscillate around 
their equilibrium position with an angular frequency of $D\sqrt{2}\omega$.
One notes that the electrons oscillate with the same angular frequency 
in both the Moshinsky atom and $D$-HL-spherium (see above).

The first-order correction of the kinetic energy
\begin{equation}
	\Hat{H}^{(1)} = \frac{D-1}{3} \theta \frac{d}{d\theta}
\end{equation}
and the zeroth-order wave function \eqref{psi-largeR} yield the asymptotic expansion \cite{Seidl07b,TEOAS1}
\begin{equation} \label{largeR-TEOAS}
	E	\simeq	\frac{D\omega}{\sqrt{2}} - \frac{D(D-1)}{6} + \ldots,
\end{equation}
which shows that the next term of this expansion is different in the Moshinsky atom and $D$-HL-spherium
than their Coulombic analogues.

For comparison, in $D$-spherium (the Coulombic analog of $D$-HL-spherium),
one can eventually show that the angular frequency is equal to $D/(2R^{3/2})$, 
and the asymptotic expansion reads \cite{LoosUnpub}
\begin{equation} \label{spherium-asy}
	E \simeq \frac{1}{2R} + \frac{D}{4R^{3/2}} - \frac{D(9D-14)}{64 R^2} + \ldots,
\end{equation}
where the first term in \eqref{spherium-asy} represents the classical mechanical energy 
of two electrons sitting on opposite sides of a sphere of radius $R$, and the third term 
is the first anharmonic correction.

\subsection{\label{subsec:3D}Exact solvability for $D=1$ and $D=3$}

\begin{figure}                  
\begin{center}
\caption{\label{fig:TEOAS-psi}$\psi_n$ as a function of $\theta$ for the unit force constant, 
and various $n$ [$n$=0 (solid), 1 (dashed), 2 (dotted) and 3 (dot-dashed)].}
        \subfigure[~$D = 1$]{
        	\includegraphics[width=0.45\textwidth]{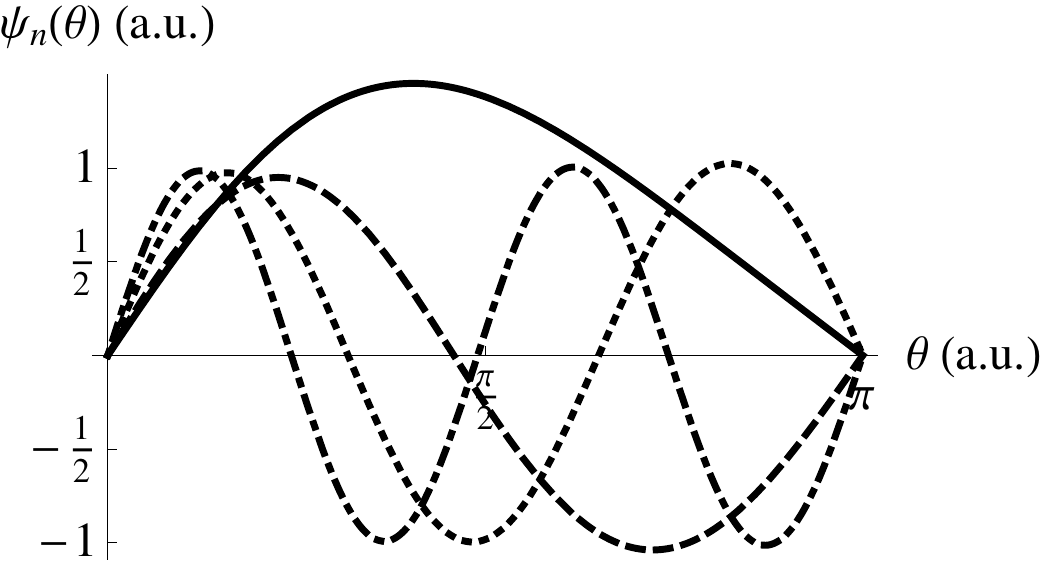}
	}
        \subfigure[~$D = 3$]{
        	\includegraphics[width=0.45\textwidth]{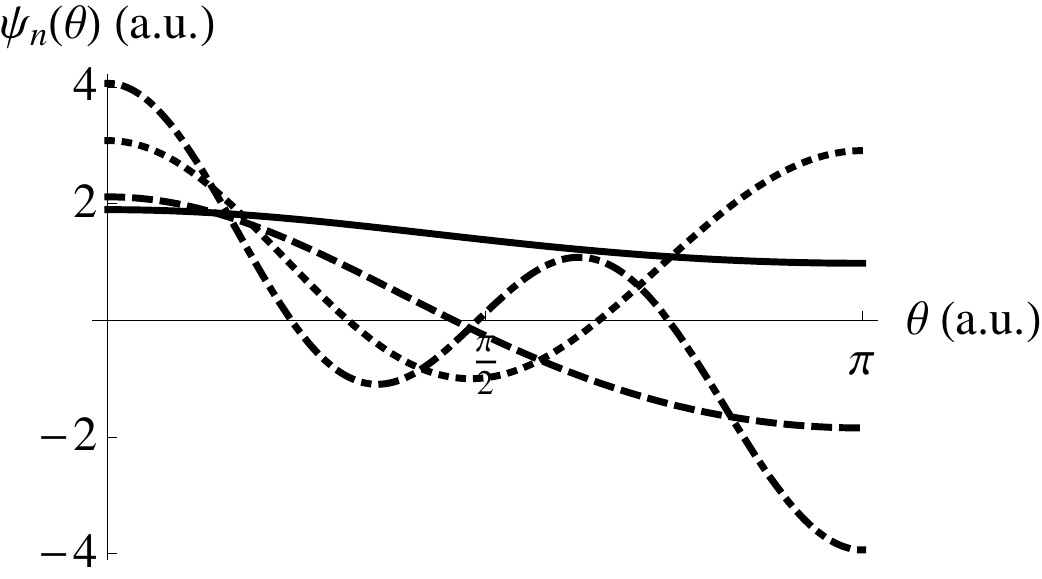}
	}
\end{center}
\end{figure}

Using the ansatz
\begin{equation}
	\psi (\theta) = \left( \sin \theta \right)^{\frac{2}{D-1}} \chi (\theta)
\end{equation}
Eq. \eqref{Hs-TEOAS} is recast as
\begin{equation} \label{Mathieu-Eq}
	\frac{d^2 \chi (\theta)}{d\theta^2} 
	+ \left[ \frac{D-1}{2} - \frac{(D-1)(D-3)}{4} \cot^2 \theta \right] \chi (\theta)	
	- \omega^2 \left(1 - \cos \theta \right) \chi (\theta) + E \chi (\theta) = 0,
\end{equation}
which coincide with the Mathieu differential equation \cite{Mathieu,ASbook} for $D = 1 \text{ and } 3$, 
due to the cancellation of the second term in brackets in \eqref{Mathieu-Eq}.

Then, the cases $D = 1$ and $D = 3$ are exactly solvable, and the exact wave functions of 
the $n$th excited state are
\begin{align}
	\psi_n^{\rm 1D} (\theta) 	&	= \Se \left( b_{2(n-1)},-2 \omega^2,\frac{\theta}{2} \right),	
						\\
	\psi_n^{\rm 3D} (\theta) 	& 	= \frac{1}{\sin \theta} 
						\Se \left( b_{2(n-1)},-2 \omega^2,\frac{\theta}{2} \right),	
\end{align}
where $\Se$ is the the odd Mathieu function, and $b_{2(n-1)}$ its characteristic value 
($n \in \mathbb{N}$ is the number of nodes in the wave function between $0$ and $\pi$) \cite{ASbook}.
The energies of the $n$th excited state are given by 
\begin{align}
	E_n^{\rm 1D} 	&	= 	\omega^2 + \frac{b_{2(n-1)}}{4},	\label{E1d} \\
	E_n^{\rm 3D} 	&	= 	\omega^2 + \frac{b_{2(n-1)}}{4} - 1,	\label{E3d}
\end{align}
The Taylor expansions of \eqref{E1d} and \eqref{E3d} demonstrate that the ground state energy behaves 
consistently with Eqs. \eqref{smallR-TEOAS} and \eqref{largeR-TEOAS} 
for small- and large-$\omega$, respectively.
Figure \ref{fig:TEOAS-psi} shows the ground state and the first three excited states 
for $D = 1$ and $D = 3$ ($\omega = 1$ in both cases).

\subsection{\label{subsec:TEOAS-HF}Hartree-Fock approximation}

Following our previous work \cite{TEOAS1,EcLimit09}, it is straightforward to show 
that, for $D \ge 2$, the HF wave function and energy are
\begin{align}
	\psi_{\rm HF} (\theta) 	& = \frac{\Gamma\left(\frac{D+1}{2}\right)}{2 \pi^{\frac{D+1}{2}}},	&
				& E_{\rm HF} = \omega^2,
\end{align}
which yields an uniform electron density over the surface of the hypersphere.

\subsection{\label{subsec:TEOAS-corr}Correlation energy}

\begin{figure}                  
\begin{center}
\caption{\label{fig:Ec-TEOAS}$-\Ec$ in $3$-HL-spherium as a function of $\omega$ (solid curve).  
The small- and large-$\omega$ expansions are also represented (dashed and dotted curves, respectively).}
        \includegraphics[width=0.45\textwidth]{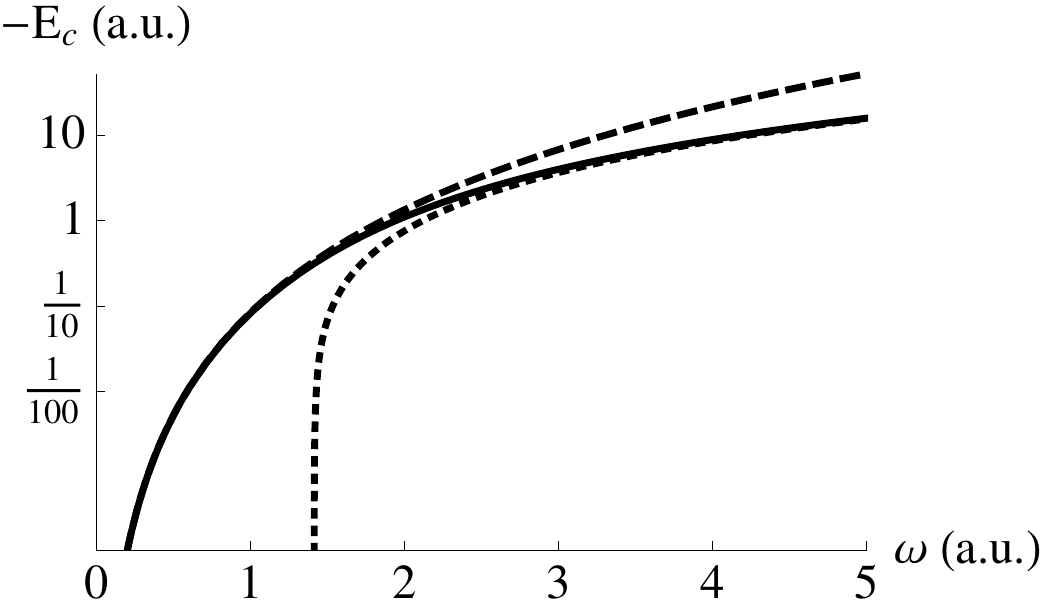}
\end{center}
\end{figure}

Following the results of the previous section, for small-$\omega$, 
the correlation energy for two electrons on a $D$-sphere behaves as
\begin{equation} \label{Ec-TEOAS-small-w}
	\Ec	\simeq	- \frac{\omega^4}{D(D+1)} + \ldots,
\end{equation}
which, like the Moshinsky atom, is quartic in $\omega$, but exhibits a different behavior with respect to $D$.
The correlation energy in 3-HL-spherium is represented in Figure \ref{fig:Ec-TEOAS}.

From Eq.~\eqref{Ec-TEOAS-small-w}, it is obvious that $\Ec$ behaves 
quadratically with respect to $D$ in the large-$D$ limit [$\Ec \simeq -\omega^4/D^2$], like $D$-spherium 
[$\Ec \simeq -1/(8D^2)$] \cite{EcLimit09}.
However, the prefactor is different.

\section{\label{sec:ccl}Conclusion}

In this article, we have studied the Hooke's law correlation energy of two model systems: 
the Moshinsky atom and the spherium model.

We have shown that, in the weakly correlated regime (small-$\omega$ limit), 
the correlation energy is quartic in $\omega$ in both systems, but behaves differently 
with respect to the dimensionality of the space. 
This feature reveals the difference between the Coulombic and the Hooke's law system.
Indeed, the correlation energy of both $D$-hookium and $D$-spherium tends to a constant 
in the high-density limit.

In the strongly correlated regime (large-$\omega$ limit) 
\cite{GoriGiorgi09b,GoriGiorgi09c}, 
the leading terms of the asymptotic expansion in the Moshinsky atom and $D$-HL-spherium are identical, 
and they represent the zero-point oscillations of the electrons when the kinetic energy tends to zero.
This could be viewed as an ``attractive'' version of the Wigner crystallization \cite{Wigner34}, 
which involves electrons in the low-density limit (quantum dots).

Moreover, we have shown that the Schr\"odinger equation of the $D$-HL-spherium reduces 
to a Mathieu differential equation \cite{Mathieu,ASbook} for two specific values of the dimension:
the model of two electrons on the surface of a sphere and interacting {\em via} a hookean potential 
is exactly solvable for $D = 1 \text{ and } 3$, and the exact wave function is based on Mathieu functions.

\begin{acknowledgments}
The author thanks Pr. Peter Gill for many stimulating discussions, 
and the Australian Research Council (Grants DP0664466 and DP0771978) for funding.
\end{acknowledgments}

\end{document}